\documentclass[a4paper,11pt]{article}
\usepackage{pos}

\preprint{CERN-TH-2026-053}

\title{$K \pi$ scattering as a step towards $B \to K^* \ell^+ \ell^-$ from Lattice QCD}

\author*[a]{Felix Erben}
\author[b]{Matthew Black}
\author[c,b]{Peter Boyle}
\author[a]{Matteo Di Carlo}
\author[b]{Vera G\"ulpers}
\author[b]{Maxwell T. Hansen}
\author[d]{Nelson Pitanga Lachini}
\author[b]{Rajnandini Mukherjee}
\author[b]{Antonin Portelli}
\author[e]{J. Tobias Tsang}

\affiliation[a]{CERN, Theoretical Physics Department, Geneva, Switzerland}

\affiliation[b]{School of Physics and Astronomy, University of Edinburgh, Edinburgh EH9 3JZ, U}

\affiliation[c]{Brookhaven National Laboratory, Upton, NY 11973, USA}

\affiliation[d]{Department of Applied Mathematics and Theoretical Physics, Center for Mathematical Sciences,\\
University of Cambridge, Wilberforce Road, Cambridge, CB3 0WA, United Kingdom}

\affiliation[e]{Theoretical Physics Division, Department of Mathematical Sciences,
University of Liverpool, Liverpool L69 3BX, UK}

\emailAdd{felix.erben@cern.ch}

\abstract{
Rare $b\to s\ell^+\ell^-$ decays provide some of the most sensitive tests of the Standard Model and require precise and systematically improvable hadronic input from lattice QCD.
For the phenomenologically important channel $B\to K^*\ell^+\ell^-$ this entails a first-principles treatment of a resonant $K\pi$ final state together with controlled heavy-quark dynamics.
We present the status of a new exploratory lattice calculation that combines a variational determination of finite-volume $K\pi$ states with the $1+J\to2$ finite-volume formalism to access the relevant matrix elements.
The computation is carried out on an RBC/UKQCD domain-wall fermion ensemble with $a^{-1} \approx 2.7\,\mathrm{GeV}$ and employs a dual heavy-quark strategy, using both a relativistic heavy-quark action tuned to the physical $b$ mass and domain-wall heavy masses extrapolating from charm.
All correlation functions are computed using (stochastic) distillation, providing a versatile setup that supports a broad range of heavy-to-light transitions into resonant final states.
We show first two-point results for the $K^*\leftrightarrow K\pi$ system and discuss the accessible kinematic region, which allows for a controlled study at high $q^2$.
The outlook for extending the calculation to lower $q^2$ and for incorporating effects from charmonium resonances is outlined.
}

\FullConference{The 42nd International Symposium on Lattice Field Theory (LATTICE2025)\\
2-8 November 2025\\
Tata Institute of Fundamental Research, Mumbai, India\\}

\begin{document}
\maketitle

\section{Introduction and motivation}
Lattice QCD has reached a level of maturity where it provides decisive, systematically improvable input for flavour physics.
At the precision frontier, prominent examples include the hadronic contributions to the muon anomalous magnetic moment~\cite{Aliberti:2025beg}
and determinations of the strong coupling $\alpha_s$~\cite{Brida:2025gii}, where lattice QCD now delivers the most precise and accurate theoretical determination.
In other applications, such as $\varepsilon_K$~\cite{Bai:2023lkr}, the ultimate target precision may be more modest, but lattice QCD plays a unique role:
it enables a first-principles and systematically controllable treatment of long-distance QCD effects that are otherwise difficult to constrain.

In these proceedings we discuss a further direction where lattice QCD may become decisive in providing stringent flavour-physics tests.
We work at the interface of precision flavour physics and multi-hadron finite-volume methods, with a programme towards lattice QCD calculations of the rare decay
$B\to K^*(\to K\pi)\,\ell^+\ell^-$.
Phenomenologically, $b\to s \ell^+\ell^-$ observables remain a key window on possible physics beyond the Standard Model.
Experimentally, the branching ratios of $B\to K\ell^+\ell^-$~\cite{LHCb:2014cxe}, $B_s\to \phi\ell^+\ell^-$~\cite{LHCb:2021zwz},
and $B\to K^*\ell^+\ell^-$~\cite{LHCb:2016ykl,CMS:2024atz,LHCb:2025mqb} have been measured with unprecedented precision, and represent one of the most promising
remaining Standard Model tests with the potential to uncover new physics.
On the theory side, relatively recent lattice QCD calculations for pseudoscalar $K$ final states exist~\cite{Parrott:2022zte,Parrott:2022rgu}.
For vector final states, the most recent lattice QCD calculation for $K^*$ and $\phi$ final states~\cite{Horgan:2013hoa} is now more than a decade old and was performed
in the narrow-width approximation, treating the hadronic final states as stable.
This is expected to be an excellent approximation for the very narrow $\phi$, but requires substantially more scrutiny for the much broader $K^*$ if one aims to draw decisive
conclusions regarding new physics.
While short-distance electroweak physics dominates many aspects of these processes, the hadronic final state introduces an essential complication:
the $K^*$ is a resonance rather than a stable particle, and a correct treatment of the $K\pi$ dynamics is required.

The theoretical ingredients required for such a programme have advanced substantially in recent years~\cite{Erben:2025zph}.
Finite-volume methods for elastic two-hadron scattering based on L\"uscher's formalism~\cite{Maiani:1990ca,Luscher:1985dn,Luscher:1986pf,Luscher:1990ux,Rummukainen:1995vs,Kim:2005gf,Fu:2011xz,Leskovec:2012gb,Hansen:2012tf,Briceno:2014oea}
are well established, and lattice determinations of $K\pi$ scattering in the $K^*$ channel exist~\cite{Fu:2012tj,Prelovsek:2013ela,Wilson:2014cna,Wilson:2019wfr,Bali:2015gji,Brett:2018jqw,Rendon:2020rtw},
including a recent study by a large part of the authors of these proceedings at the physical point~\cite{Boyle:2024hvv,Boyle:2024grr}.
Moreover, a general formalism exists to relate finite-volume matrix elements for $1+J\to 2$ transitions~\cite{Briceno:2021xlc} to the corresponding infinite-volume amplitudes,
providing the analogue of the Lellouch--L\"uscher factor for electroweak transitions into two-particle states.
Prior to the formulation of this general framework, computations of $1+J\to 2$ transitions were already carried out in concrete applications,
notably for photoproduction processes~\cite{Briceno:2016kkp,Alexandrou:2018jbt,Radhakrishnan:2022ubg}.
More recently, an exploratory application to heavy-meson decays into resonances, $B\to\rho\,\ell\nu$~\cite{Leskovec:2025gsw},
was the first to include a heavy $B$ meson in the initial state, thereby confronting an additional lattice challenge shared with the project at the centre of these proceedings.
Against this backdrop, $B\to K^*\ell^+\ell^-$ combines (i) a resonant multi-hadron final state and (ii) heavy--light dynamics involving the $b$ quark,
which together define a challenging but timely lattice target.

A central practical challenge is that simulations with physical light quarks require large volumes ($M_\pi L\gtrsim 4$) to suppress finite-volume effects,
while controlling discretization effects for the heavy quark requires small lattice spacings ($a m_h \ll 1$ for fully relativistic treatments).
Very fine lattices at large physical volume are therefore numerically expensive and may face additional algorithmic challenges such as topological freezing.
In this work we outline our strategy and present first status results from a new exploratory computation,
which leverages the versatility of (stochastic) distillation~\cite{Peardon:2009gh,Morningstar:2011ka} to cover a broad range of decay channels within a unified dataset.

\section{From form factors to finite-volume matrix elements}

The hadronic matrix elements of the $b\to s$ vector/axial and tensor currents,
$J_{V(A)}^\mu=\bar{s}\gamma^\mu(\gamma_5)b$ and $J_{T(AT)}^\mu=\bar{s}\sigma^{\mu\nu}q_\nu(\gamma_5)b$,
are parameterized in terms of seven independent form factors,
$V$, $A_{0,1,2}$ and $T_{1,2,3}$, which are functions of the momentum transfer $q^2$.
They are defined through
\begin{align}
\langle \bar{V}(k,\eta) | J_V^\mu | \bar{B}(p) \rangle
&=
\epsilon^{\mu \nu \rho \sigma} \eta^*_\nu p_\rho k_\sigma
\frac{2 V(q^2)}{M_B+M_V}\,, \\[4pt]
\langle \bar{V}(k,\eta) | J_A^\mu | \bar{B}(p) \rangle
&=
i \eta^*_\nu
\begin{aligned}[t]
\Big[
&g^{\mu \nu} (M_B + M_V) A_1(q^2)  + (p+k)^\mu q^\nu \frac{A_2(q^2)}{M_B+M_V}  \\
&-2 M_V \frac{q^\mu q^\nu}{q^2}\big(A_3(q^2)-A_0(q^2)\big)
\Big] ,
\end{aligned}
\\[4pt]
\langle \bar{V}(k,\eta) | J_T^\mu | \bar{B}(p) \rangle
&=
\epsilon^{\mu \nu \rho \sigma} \eta^*_\nu p_\rho k_\sigma
\, 2 T_1(q^2)\,, \\[4pt]
\langle \bar{V}(k,\eta) | J_{AT}^\mu | \bar{B}(p) \rangle
&=
i \eta^*_\nu
\begin{aligned}[t]
\Big[
&\Big(g^{\mu \nu} (M_B^2 - M_V^2) - (p+k)^\mu q^\nu\Big) T_2(q^2) \\
&- q^\nu
\left(
q^\mu - \frac{q^2}{M_B^2-M_V^2}(p+k)^\mu
\right) T_3(q^2)
\Big] .
\end{aligned}
\end{align}

Here $p^\mu$ and $k^\mu$ denote the four-momenta of the $B$ and vector meson respectively,
$\eta^\mu$ is the polarization vector of the final-state vector meson,
and the momentum transfer is defined as
\begin{equation}
q^\mu = p^\mu - k^\mu , \qquad q^2 = (p-k)^2 .
\end{equation}
The form factor $A_3$ is a linear combination of $A_1$ and $A_2$ and is not independent.
The tensor current matrix elements can be decomposed into three independent form factors,
which we write in terms of the vector and axial combinations
$J_T^\mu$ and $J_{AT}^\mu$ for convenience.

A lattice determination of these form factors provides the nonperturbative input required for precision Standard Model predictions of
$b\to s\ell^+\ell^-$ observables.

For an unstable final state the calculation proceeds through finite-volume matrix elements.
The first step is the construction of interpolating operators with the quantum numbers of the $K^*$ in the relevant kinematics.
We employ a variational basis including both quark--antiquark and two-hadron operators,
\begin{equation}
  O_i \in \{\, V,\; K_1\pi_1,\; K_2\pi_2,\; \dots \,\},
\end{equation}
where $V$ denotes the vector bilinear with total momentum $P$ and
$K_i\pi_i$ represents a $K\pi$ interpolator with total momentum
$p_\pi+p_K=P$ corresponding to the $i^{\mathrm{th}}$ lowest-energy
two-particle state. From this basis, we form a matrix of correlation functions
\begin{equation}
  C_{ij}(t)=\langle O_i(t)\, O_j^\dagger(0)\rangle .
\end{equation}
Solving the generalized eigenvalue problem~\cite{Michael:1982gb,Michael:1985ne,Luscher:1990ck,Blossier:2009kd} yields eigenvectors $v^{(n)}$ that define optimized interpolators for the finite-volume energy eigenstates in the $K\pi$ channel.

These eigenvectors are then used to project three-point correlation functions,
\begin{equation}
\langle O_{K^*}(t)\, J^\mu(t_J)\, O_B^\dagger(t_B)\rangle
=
\sum_i v_i^{(n_{K^*})}
\langle O_i(t)\, J^\mu(t_J)\, O_B^\dagger(t_B)\rangle ,
\end{equation}
which gives access to the finite-volume matrix elements
\begin{equation}
  \langle n;\mathbf{k} | J^\mu_{V/A/T} | B,\mathbf{p}\rangle_L ,
\end{equation}
where $|n;\mathbf{p}_{K\pi}\rangle$ denotes a discrete finite-volume eigenstate.

The relation to the infinite-volume amplitudes is obtained using the
$1+J\to2$ finite-volume formalism~\cite{Briceno:2021xlc}.
This introduces a finite-volume form factor that is mapped to the physical form factor
through an analogue of the Lellouch--L\"uscher factor, which depends on the
$K\pi$ scattering amplitude in the relevant partial waves.
In the narrow-width limit the familiar stable-particle form-factor description is recovered,
while the full formalism enables a systematically controlled treatment of the resonant channel.

\section{Exploratory computation: ensemble and heavy-quark strategy}

We present first status from an exploratory study based on the RBC/UKQCD domain-wall fermion ensemble F1M~\cite{Boyle:2017jwu,Boyle:2018knm},
with inverse lattice spacing $a^{-1} \approx 2.7~\mathrm{GeV}$ and light/strange pseudoscalar masses
\begin{equation}
  M_\pi = 232~\mathrm{MeV}\,,\qquad M_K = 510~\mathrm{MeV}\,,
\end{equation}
on a $48^3\times96$ lattice.
For the $K\pi$ system we include all momenta satisfying
\(
0 \leq \left(\tfrac{L}{2\pi}\right)^2 \mathbf{k}^2 \leq 4
\),
for the individual kaon, pion and total $K\pi$ momentum.
For the $B$ meson we use $\mathbf{p}=[000]$ and $[111]$.

A key challenge for $B$-physics calculations is the simultaneous control of finite-volume effects,
which require large values of $M_\pi L$, and heavy-quark discretization effects, which scale with $a m_h$.
Our strategy combines two complementary heavy-quark approaches on the same ensemble:
\begin{itemize}
  \item a relativistic heavy-quark (RHQ) action~\cite{Christ:2006us,Lin:2006ur} tuned directly to the physical $b$-quark mass;
  \item three additional domain-wall heavy masses in the range
$m_c \le m_h \lesssim 0.5\,m_b$,
providing a lever arm in the heavy-quark mass and enabling an exploratory interpolation between charm and bottom.
\end{itemize}
This approach yields both a direct determination at $m_b$ and a lever arm in the heavy-quark mass,
and is designed to support a wider programme of heavy-to-light transitions.

\begin{figure}[t]
  \centering
  \includegraphics[width=0.62\textwidth]{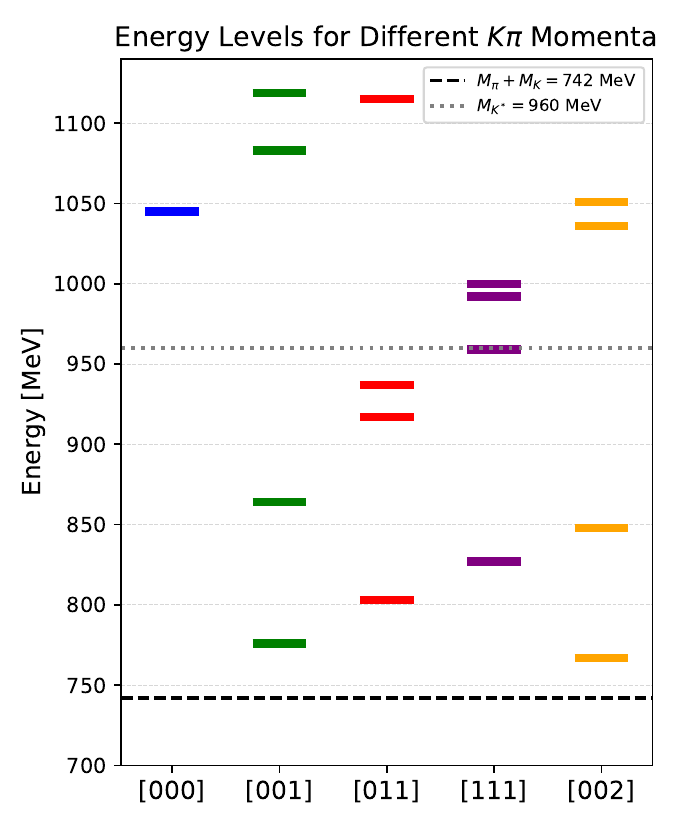}
  \caption{Expected finite-volume energy levels (free spectrum) in the $K\pi$ channel on the F1M ensemble around the $K^*$ region, shown relative to the nominal resonance mass $M_{K^*}$. The resonance region lies well above the $K\pi$ threshold: on F1M we have $M_{K^*}\simeq 960~\mathrm{MeV}$ and $M_\pi+M_K\simeq 742~\mathrm{MeV}$.}
  \label{fig:F1M_levels}
\end{figure}

The kinematic window accessible in this setup is illustrated in Fig.~\ref{fig:F1M_levels},
which shows the free finite-volume energy levels in the $K\pi$ channel on the F1M ensemble
in the vicinity of the expected $K^*$ resonance mass.
The resonance region is well separated from the $K\pi$ threshold
($M_{K^*}\simeq960~\mathrm{MeV}$ compared to $M_\pi+M_K\simeq742~\mathrm{MeV}$),
and the relevant energy range can therefore be mapped with good resolution,
providing a favorable regime for controlling heavy-quark discretization effects.

\section{Distillation for electroweak transitions into resonances}

All two- and three-point correlation functions are computed using distillation~\cite{Peardon:2009gh,Morningstar:2011ka},
which projects the quark fields onto a low-rank Laplacian–Heaviside (LapH) subspace,
\begin{equation}
  \mathcal{S}(t)=V^\dagger(t)\,V(t)\,,
\end{equation}
where $V(t)$ is the matrix whose columns are the lowest eigenvectors of the three-dimensional gauge-covariant lattice Laplacian on timeslice $t$.
This enables an efficient construction of extended operator bases,
including two-hadron interpolators, together with the required multi-hadron contractions.
For three-point functions with local electroweak currents, an unsmeared current insertion is required~\cite{Mastropas:2014fsa}.
Using $\gamma_5$-hermiticity, the quark-line contractions can be arranged such that one current insertion remains fully local.

We employ a hybrid setup with stochastic distillation for sink-to-sink quark lines and exact distillation elsewhere.
The parameters of the exploratory calculation are:
\begin{itemize}
  \item $N_v = 60$ LapH eigenvectors,
  \item exact distillation solves on every fourth timeslice (24 out of 96),
  \item stochastic distillation with dilution scheme $(LI,TI,SF)=(8,16,4)$ and 6 noise sources, using the notation of~\cite{Morningstar:2011ka}.
\end{itemize}

The stochastic quark lines provide estimators for all two-point functions at all time separations.
For three-point functions the local current insertion time can be varied freely,
while the $B$ and $K\pi$ interpolating operators are restricted to the timeslices where exact distillation solves are available.
We explore source–sink separations in the range
\(
12 \leq \Delta T /a \leq 36
\),
chosen to suppress excited-state contamination while retaining a statistically significant signal.

A practical consequence of the local current insertion is a reduced I/O efficiency compared to standard distillation:
the unsmeared vectors cannot be absorbed into compact perambulators and intermediate objects must be buffered.
In the present setup this requires temporary storage at the level of $\mathcal{O}(500~\mathrm{TB})$
for all heavy- and light-quark current-transition meson fields.
The payoff is the broad physics reach of a unified dataset,
enabling $b\to s$, $b\to d$, $b\to u$ as well as $c\to s$, $c\to d$ and $c\to u$ electroweak transitions into resonant final states.

All software used in this project is based on Grid~\cite{Boyle2015} and Hadrons~\cite{Hadrons2023},
with details of the distillation implementation and its application to scattering given in~\cite{Lachini:2024ugs}.

\section{First status: $K^*\leftrightarrow K\pi$ two-point functions}

As a first step we analyse two-point correlation functions in the $K^*$ channel including $K\pi$ interpolating operators.
In the rest frame this leads to a symmetrized $5\times5$ correlation matrix,
from which we inspect the signal quality prior to a full GEVP analysis.

At the time of writing, data from 8 gauge-field configurations are available,
with 24 source timeslices per configuration.
For this preliminary study, the 24 source positions per configuration are grouped into three bins, giving three measurements per configuration that are treated as statistically independent. The resulting uncertainties are therefore expected to be underestimated.
The target statistics for the exploratory calculation are $30$--$60$ configurations,
after which a stable GEVP analysis and the extraction of the finite-volume spectrum will be carried out.

\begin{figure}[t]
  \centering
  \includegraphics[width=\textwidth]{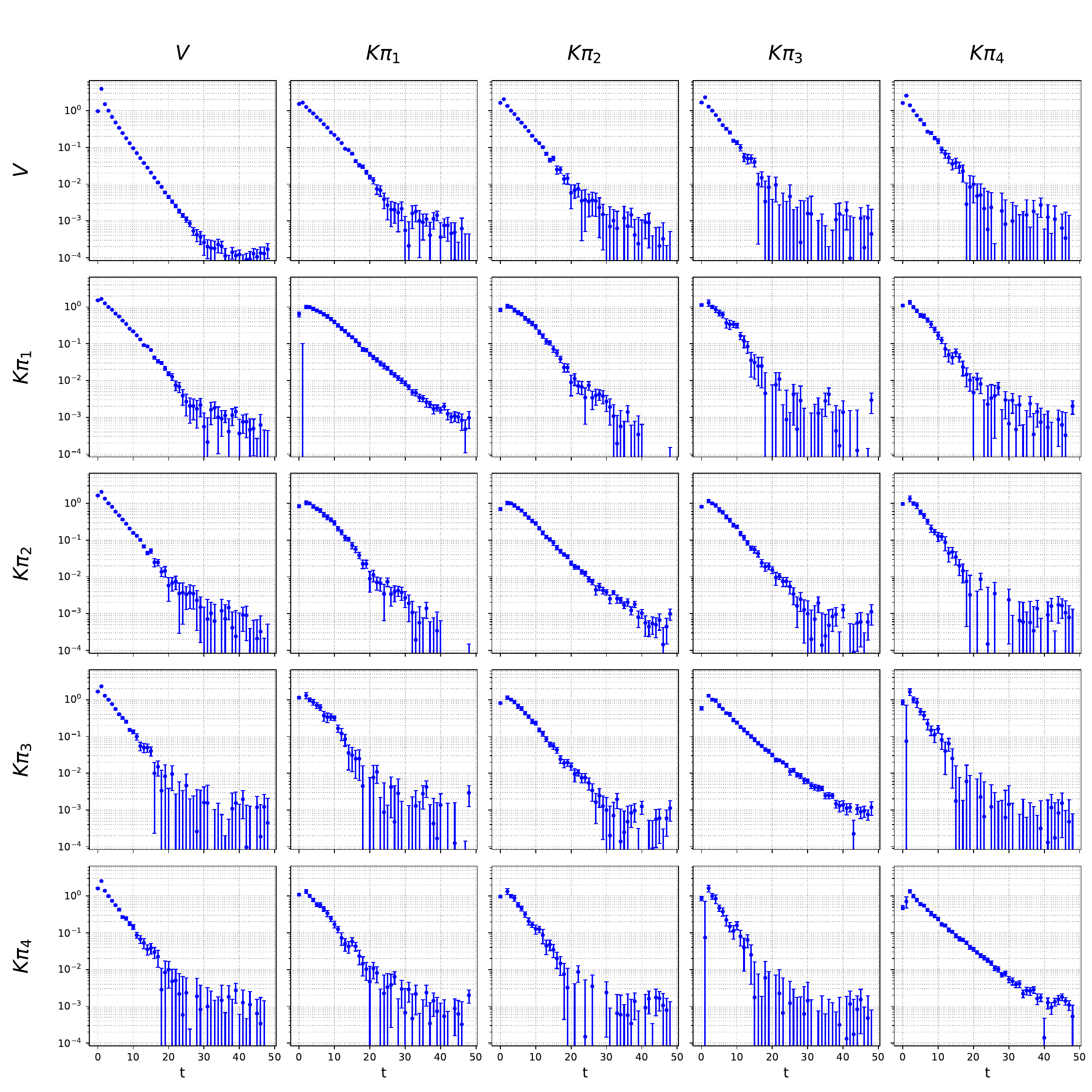}
  \caption{Matrix of two-point correlation functions entering the $5\times5$ GEVP in the $K^*$ channel in the rest frame on the F1M ensemble. 
Each panel corresponds to one element $C_{ij}(t)$ of the correlation matrix built from the operator basis 
$\{V,\,K_1\pi_1,\,K_2\pi_2,\,K_3\pi_3,\,K_4\pi_4\}$.
All correlators are normalized to the $V$–$V$ two-point function so that a common vertical scale can be used across panels.
The data shown are based on 8 gauge configurations with 24 source times per configuration; 
for this exploratory look three bins per configuration are treated as statistically independent, so the uncertainties are expected to be underestimated.}
  \label{fig:F1M_corr}
\end{figure}

A first look at the correlation matrix is shown in Fig.~\ref{fig:F1M_corr}.
Each panel corresponds to one element of the $5\times5$ matrix entering the GEVP in the rest frame.
At this stage the figure serves purely as a qualitative demonstration of the data quality and operator basis,
while a quantitative analysis will require the full target statistics.

\section{Kinematic considerations and outlook}

Lattice calculations of heavy-to-light form factors are most precise near maximum momentum transfer,
$q^2_{\max}$, where $\mathbf{p}=\mathbf{k}=0$.
Reaching low $q^2$ requires large recoil of the hadronic system, i.e.\ large $|\mathbf{k}|$.
On the present $48^3\times96$ lattice with $a^{-1}=2.7~\mathrm{GeV}$,
achieving $q^2=0$ would require momenta of order
$\mathbf{k}\sim[8,8,8]$ in units of $2\pi/L$,
which is not feasible in practice.
In addition, at such energies the $K\pi$ system encounters multiple inelastic thresholds,
beyond the regime where the elastic finite-volume formalism for $K\pi$ scattering and $1+J\to2$ transitions applies.
For these reasons the near-term emphasis of this study is on the high-$q^2$ region,
where lattice control is strongest.

Access to lower $q^2$ can be achieved on coarser lattices,
where larger spatial momenta are available at fixed integer Fourier modes,
at the price of a reduced reach towards the physical $b$-quark mass in a fully relativistic setup.
With a sufficiently large set of ensembles, both limits can be controlled through a combined continuum and heavy-quark extrapolation,
in close analogy to the strategy successfully employed for $B\to K\ell^+\ell^-$~\cite{Parrott:2022zte,Parrott:2022rgu}.

A further challenge arises from the presence of charmonium resonances in the intermediate-$q^2$ region.
Experimentally, their impact on neighbouring kinematic regions has been found to be larger than early theory estimates anticipated~\cite{LHCb:2013ywr}.
A fully controlled lattice calculation of $B\to K^*$ form factors, in particular if extended towards low $q^2$,
will therefore require a quantitative treatment of these effects.
First progress in this direction has recently been achieved using spectral-reconstruction techniques~\cite{Frezzotti:2025hif}
in the context of $B\to K\ell^+\ell^-$ and $\bar{B}_s\to\ell^+\ell^-\gamma$.

In summary, we have initiated a new exploratory lattice study of heavy-meson decays into resonant final states,
with $B\to K^*(\to K\pi)\,\ell^+\ell^-$ as a key application.
The calculation combines a variational treatment of the $K\pi$ final state,
a dual heavy-quark strategy (RHQ at $m_b$ together with domain-wall heavy masses interpolating from charm),
and a versatile distillation setup that supports a wide range of electroweak transitions.
First two-point data in the $K^*$ channel are available and three-point functions are under analysis.
The immediate next steps are to increase the statistics to the target $40$--$60$ configurations,
stabilize the GEVP extraction of the finite-volume states,
and incorporate the $K\pi$ scattering input required to map the finite-volume matrix elements to physical amplitudes.

\section*{Acknowledgements}
F.E. has received funding from the European Union's Horizon Europe research and innovation programme
under the Marie Sk\l{}odowska-Curie grant agreement No.\ 101106913. M.D.C. has received funding from the European Union’s Horizon Europe research and innovation programme under the Marie Sk\l{}odowska-Curie grant agreement No.\ 101108006. M.B., V.G., M.T.H., R.M. and A.P. are supported in part by UK STFC grants ST/X000494/1 and ST/T000600/1. N.P.L. acknowledges support from the U.K. Science and Technology Facilities Council (STFC) [grant numbers ST/T000694/1, ST/X000664/1]. This work used the DiRAC Extreme Scaling service (Tursa) at the University of Edinburgh, managed by the EPCC on behalf of the STFC DiRAC HPC Facility (www.dirac.ac.uk). The DiRAC service at Edinburgh was funded by BEIS, UKRI and STFC capital funding and STFC operations grants. DiRAC is part of the UKRI Digital Research Infrastructure.


\begingroup
\small
\setlength{\itemsep}{0pt}
\setlength{\parskip}{0pt}
\setlength{\parsep}{0pt}
\setlength{\topsep}{0pt}
\setlength{\partopsep}{0pt}
\setlength{\bibsep}{4.5pt}
\linespread{0.92}\selectfont
\bibliographystyle{JHEP}
\bibliography{proceedings}
\endgroup

\end{document}